\documentclass[abstract=on,notitlepage,superscriptaddress,nofootinbib,aps,preprint,prd]{revtex4}
\usepackage{srcltx}
\usepackage{epsf}
\usepackage{amsthm}
\usepackage{amsfonts}
\usepackage{amsmath}
\usepackage{mathrsfs}
\usepackage{epsfig}
\usepackage{enumerate}
\usepackage{lipsum}
\usepackage{graphicx}
\usepackage{epsf}
\usepackage{subfig}
\usepackage{color}
\usepackage{caption}
\usepackage{subfig}
\usepackage[dvipsnames]{xcolor}
\usepackage{epstopdf}
\usepackage{float}
\usepackage{dcolumn}
\usepackage{hyperref}
\usepackage{amsthm}
\usepackage{graphicx, amsmath, amssymb, bm}

\newtheorem{proposition}{Proposition}

\newcommand{\be}{\begin{equation}}
\newcommand{\ee}{\end{equation}} 
\newcommand{\eei}{\end{equation}\indent\indent}
\newcommand{\bc}{\begin{center}}
\newcommand{\ec}{\end{center}}
\newcommand{\ber}{\begin{eqnarray*}}
\newcommand{\ear}{\end{eqnarray*}}
\newcommand{\ba}{\begin{array}}
\newcommand{\ea}{\end{array}}

\newcommand{\bea}{\begin{eqnarray}}
\newcommand{\eea}{\end{eqnarray}}

\newcommand{\ei}{\end{itemize}}

\begin{document}


\title{The Buchdahl Bound Denotes The Geometrical Virial Theorem}



\author{Naresh Dadhich}\email[]{nkd@iucaa.in}
\affiliation{Inter-University Centre for Astronomy and Astrophysics, Post Bag 4, Pune, 411 007, India}
\affiliation{Astrophysics Research Centre, School of Mathematics, Statistics and Computer Science, University of KwaZulu--Natal, Private Bag X54001, Durban 4000, South Africa}
\author{Rituparno Goswami}\email[]{goswami@ukzn.ac.za}
\affiliation{Astrophysics Research Centre, School of Mathematics, Statistics and Computer Science, University of KwaZulu--Natal, Private Bag X54001, Durban 4000, South Africa}
\author{Chevarra Hansraj}\email[]{chevarrahansraj@gmail.com}
\affiliation{Astrophysics Research Centre, School of Mathematics, Statistics and Computer Science, University of KwaZulu--Natal, Private Bag X54001, Durban 4000, South Africa}


\begin{abstract}
In this paper, we geometrically establish yet another correspondence between Newtonian mechanics and general relativity by connecting the Buchdahl bound and the Virial theorem. Buchdahl stars are defined by the saturation of the Buchdahl bound, $\Phi(R)\leq4/9$ where $\Phi(R)$ is the gravitational potential felt by a radially falling particle. An interesting alternative characterization is given by gravitational energy being half of non-gravitational energy. With insightful identification of the former with kinetic and the latter with potential energy, it has been recently argued that the equilibrium of a Buchdahl star may be governed by the Virial theorem. In this paper, we provide a purely geometric version of this theorem and thereby of the Buchdahl star characterization. We show that the condition for an accreting Buchdahl star to remain in the state of Virial equilibrium is that it must expel energy via heat flux, appearing in the exterior as Vaidya radiation. If that happens then a Buchdahl star continues in the Virial equilibrium state without ever turning into a black hole.
 
\end{abstract}


\maketitle


\section{Introduction}

Differentiable properties, smoothness and the regularity of field equations for a wide class of gravitational theories lead to a number of bounds on the structure of compact objects (for example, see \cite{chandra} for the case of general relativity). One of these interesting upper bounds in general relativity is the Buchdahl bound \cite{buch1} which was obtained under very general conditions of the fluid pressure being isotropic, density decreasing outwards and matching the metric at the boundary to the Schwarzschild vacuum metric.

Subsequently, there have been several alternative derivations of the bound under different conditions and assumptions \cite{buch2, bondi, islam, wald, stephani, and1, kar} for spherical and static compact stars adopting the Schwarzschild exterior vacuum spacetime. This bound states that the mass $(\mathcal{M})$ to radius $(R)$ ratio must be
\begin{equation}
\frac{ \mathcal{M}}{R} \leq \frac49,
\end{equation}
 of any regular, thermodynamically stable perfect fluid star. Since it is a weaker compactness bound in comparison to the black hole (the Schwarzschild static limit of $\mathcal{M}/R =1/2$), it indicates that even though the compact object is not trapped, a stable static spherical configuration cannot occur for $4/9 < \mathcal{M}/R < 1/2$ \cite{buch1}. In section \ref{sect2}, we give a mathematical definition of the Buchdahl theorem and its characterizations.
 
This bound is established using the regularity and smoothness of metric functions in the compact stellar interior and the Israel Darmois matching conditions at the boundary of the star where the spacetime is smoothly matched to a Schwarzschild exterior via the matching of the first and second fundamental forms. Additionally, imposing the condition of thermodynamic stability will imply the average density should be monotonic and non-increasing from the centre to the surface of the star. An important feature of this bound is that it is model independent. Hence it is true for any physically realistic equation of state for the stellar matter. A direct consequence of this bound is that the gravitational redshift $z$ at the stellar surface is bounded from above ($z\leq2$) \cite{wald,stephani}.

Most remarkably it was observed some time back \cite{jcap1} that the Buchdahl bound could be expressed as the ratio of gravitational ($E_{G}$) and non-gravitational ($E_{NG}$) energy as
\begin{equation} \label{half}
\frac{E_{G}}{E_{NG}}\leq \frac12.
\end{equation}
This tells us that the degree of compactness of a collapsing star is governed by the gravitational energy it has gained in attaining the radius it has now acquired. The gravitational energy could be computed by using the prescription of Brown-York quasi-local energy \cite{bro-yor} giving the total energy contained inside radius $R$. By subtracting $E_{NG}$, the mass at infinity from it, we obtain $E_{G}$ which for large $R$ attains the expected value $\mathcal{M}^2/2R$. Interestingly, all this could be done by employing the unique exterior metric without reference to the interior at all. It is very remarkable that the compactness bound could be found solely from the unique exterior metric as it illuminates the role gravitational energy plays in determining it.

Further on the saturation of \eqref{half}, this is indeed the equilibrium condition for the Buchdahl star where gravitational attraction due to mass (non-gravitational energy) is counterbalanced by the internal energy which is equal to the gravitational energy lying outside the star. This is because the Brown-York quasilocal energy prescription \cite{bro-yor} gives the total energy contained inside the star which includes bare mass $\mathcal{M}$ when it was in an infinitely dispersed state plus the measure of gravitational energy that lies outside. This is how the measure of gravitational energy is equal to the internal energy. When the former is half of the latter, it characterizes the equilibrium for the Buchdahl star. In particular, if one envisages that in the limiting state of compactness, the star interior may consist only of particles in motion interacting only through gravity then internal energy would be the kinetic energy. That would then imply that the Buchdahl star equilibrium is indeed governed by the Virial-like relation. That is to say, {\em the Buchdahl star must be a Virial star} \cite{jcap2}. The Buchdahl star defining condition may be looked upon as the relativistic generalization of the Virial theorem indicating the equilibrium of its interior.

If we were to consider the most compact equation of state, we would naturally be led to the incompressible fluid distribution where the density is constant. There the saturation of the Buchdahl bound will occur for the pressure being infinite at the center. One can therefore argue that the Buchdahl bound cannot be saturated for any fluid distribution. Hence a purely gravitational system with free particles moving is required. This concept is important as the Virial theorem is applicable for a system of particles in motion and interacting only through gravity. Such a situation is offered by the Vlasov kinetic matter governed by the Einstein-Vlasov equations  \cite{And11, And-Rein06}. Andreasson \textit{et al} \cite{And11, And-Rein06} have carried out a comprehensive investigation of stellar interior models with the Vlasov kinetic matter (particles in motion interacting only through gravity) as the source. They find that as the Buchdahl bound saturates, the interior tends to be a thin shell satisfying the strong energy condition, $p_r/\rho \to 0, 2p_t/\rho \to 1$.   

The paper is organized as follows: In section \ref{sect2}, we define the Buchdahl star and explain how its equilibrium is governed by the Virial theorem. In section \ref{sect3}, a brief overview of the $1+1+2$ covariant formalism is provided along with definitions relating to important geometrical quantities and their derivatives. The kinetic energy of the particles of the system and the total average potential energy are expressed according to geometrical quantities in section \ref{sect4}. It is here that we geometrically establish yet another correspondence between Newtonian mechanics and general relativity via connecting the Buchdahl bound and the Virial theorem. In literature, there exists many ways to link Newtonian mechanics and general relativity but this is a novel connection and result. In section \ref{sect6}, we consider an example of a continuously accreting compact star in the state of Virial equilibrium to determine if it is possible to continue in that state. We obtain an explicit expression for the heat flux if the star continues to evolve as a Buchdahl star. Then, we match the interior spacetime with an outgoing Vaidya exterior spacetime as the heat flux must be radiated out for a collapsing star and comment on its eventual fate. We use units $c = G = 1$ in the calculations of this paper. We discuss our results in section \ref{sect7}.

\section{Buchdahl star and Virial theorem} \label{sect2}

In astrophysics, Buchdahl's theorem \cite{buch1} sets a limit on the maximum possible mass that could be confined in a given radius for a spherically symmetric static star.  This bound states that:\\
{\it If a static spherically symmetric perfect fluid matter distribution has the Schwarzschild vacuum as the exterior, and the matter distribution is thermodynamically stable (that is the matter density is a non-increasing function of the radius), then the mass to radius ratio $\mathcal{M}/R$ must be less than or equal to $4/9$.}\\
It is important to note that this result is extremely robust and does not depend on the equation of state. The Buchdahl star is defined when the Buchdahl bound is saturated \cite{maxforce}, that is
\begin{equation}
\frac{\mathcal{M}}{R} =\frac49.
\end{equation} 
It is characterized, in general, irrespective of the star being neutral or charged by $\Phi(R)=4/9$, where $\Phi(R)$ is the potential felt by a radially falling test particle. On the other hand, a black hole is characterized by $\Phi(R)=1/2$. A black hole with a null boundary is naturally the most compact object. On the other hand, the Buchdahl star is the most compact non-horizon object with a timelike boundary.

Recently it has been argued by one of us (ND) \cite{jcap2} that the Buchdahl star is indeed a Virial star in the sense that its equilibrium is governed by the Virial theorem. According to the Brown-York quasi-local energy prescription \cite{bro-yor}, as the star collapses its energy in the interior increases by the amount equal to $E_{G}$ lying exterior to the star. This manifests as internal energy, which in the particular case of a Vlasov-like kinetic matter distribution, comprising of free particles in angular motion interacting only through gravity, would be a measure of average kinetic energy. On the other hand, $E_{NG}$ would be a measure of average potential energy. Then the defining relation
\begin{equation}
E_{G}=\frac{1}{2}E_{NG},
\end{equation} 
for the Buchdahl star, is indeed the Virial relation where average kinetic energy $\langle T \rangle$ is equal to half of the average potential energy $\langle V \rangle$. Its equilibrium is therefore governed by the Virial theorem and hence it is a Virial star. This discovery is the trigger for this paper, necessitating the present consideration of gaining clearer geometric insight into this new and novel perspective.

\section{Spherically symmetric spacetimes} \label{sect3}

\subsection{LRS-II spacetime decomposition}
To put the Buchdahl bound in a proper geometric perspective, we recall that spherically symmetric spacetimes are a subclass of non-rotating locally rotationally symmetric (LRS) spacetimes. Hence, they are equipped with a preferred spatial direction that is covariantly defined at every point. Using the timelike unit vector $u^a$ (defined along the fluid flow lines) and the unit vector along the preferred spatial direction $e^a$ (orthogonal to $u^a$), we can decompose the spacetime \cite{clarksonbarrett, clarkson2007, goswamiellis} as
\begin{equation}\label{decomp}
g_{ab} = -u_au_b+e_ae_b+N_{ab}\, ,
\end{equation}
where $N_{ab}$ is the 2-dimensional metric on the spherical 2-shell. Four important derivatives are introduced at this point. They are the fully orthogonally projected covariant spatial derivative `\,${D}$\,'; the covariant time derivative defined by the timelike congruence and denoted by `\,${^{\cdot}}$\,'; the covariant spatial derivative defined by the spatial congruence and denoted by `$\,\, \hat{}\,\, $'; and the $\delta$-derivative which is the projected spatial derivative onto the 2-sheet. Their definitions are given by
\begin{eqnarray}
D_{e} Z_{a...b}{}^{c...d} &=& h^r{}_{e} h^p{}_{a}... h^q{}_{b} h^c{}_{f}... h^d{}_{g}\nabla_{r} Z^{f...g}{}_{p...q}, \\
\dot{Z}_{a ... b}{}^{c ... d} &=& u^{e}\nabla_{e} Z_{a ... b}{}^{c ... d}, \\
\hat{Z}_{a...b}{}^{c...d} &\equiv& e^{f} D_{f} Z_{a...b}{}^{c...d}, \\
\delta_{f}Z_{a...b}{}^{c...d} &\equiv& N_{f}{}^{j} N_{a}{}^{l} ... N_{b}{}^{g} N_{h}{}^{c} ... N_{i}{}^{d}  D_{j}Z_{l...g}{}^{h...i}, 
\end{eqnarray}
for any tensor ${Z_{a...b}{}^{c...d}}$.

The geometrical quantities defined for the timelike congruence are the expansion scalar $(\Theta = D_{a} u^{a} )$, acceleration 3-vector $(\dot{u}^a = u^{b}\nabla_{b}u^a )$ and the shear 3-tensor $\left[\sigma_{ab} = \left(h^{c}{}_{(a} h^{d}{}_{b)} - \frac{1}{3}h_{ab} h^{cd}\right)D_{c} u_{d} \right]$. The timelike congruence uniquely defines the electric part of the Weyl tensor $(E_{ab} = C_{acbd} u^{c} u^{d} = E_{<ab>})$, whereas the magnetic part vanishes identically due to spherical symmetry. Angle brackets represent the projected, symmetric and trace-free part of tensors. This congruence can further decompose the energy momentum tensor of the matter to give the energy density $(\mu = T_{ab} u^{a} u^{b})$, isotropic pressure  $(p = \frac{1}{3} h_{ab} T^{ab})$, heat flux 3-vector $(q_a = q_{<a>} = - h^{c}{}_{a} T_{cd} u^{d})$ and anisotropic stress 3-tensor $(\pi_{ab} = T_{cd} h^c{}_{<a} h^d{}_{b>})$. The only non-vanishing geometrical quantity related to the preferred spacelike congruence is the volume expansion $(\phi = \delta_a e^a)$. Using this preferred spatial congruence, we can then extract the set of covariant scalars that govern the dynamics of the system in the following way
\begin{equation}
\mathcal{A} =\dot{u}^ae_a, \quad \Sigma=\sigma_{ab}e^ae^b, \quad \mathcal{E}=E_{ab}e^ae^b, \quad Q=q^ae_a, \quad \Pi=\pi_{ab}e^ae^b.
\end{equation}
Therefore, the set of quantities that fully describe the spherically symmetric class of spacetimes are 
\begin{equation}
\left(\Theta, \,\, \mathcal{A}, \,\, \Sigma, \,\, \mathcal{E}, \,\, \phi,  \,\, \mu, \,\, p, \,\, \Pi, \,\, Q\right).
\end{equation}
These quantities, together with their directional derivatives along $u^a$ (denoted by a dot) and $e^a$ (denoted by a hat) completely specify the corresponding Ricci and Bianchi identities of these vectors and thereby specify the dynamics completely. The complete set of the associated field equations is found in \cite{clarkson2007}.

\subsection{Gaussian curvature}

The geometry of the 2-dimensional spherical shell, that is spanned by $N_{ab}$, is completely determined by the Gaussian curvature which is the inverse square of its area radius.
In terms of the $1+1+2$ covariant scalars, we can write the Gaussian curvature of the 2-shell as
\begin{equation}
\label{K}
K = \frac{1}{3}\mu - \mathcal{E} - \frac{1}{2}\Pi + \frac{1}{4}\phi^2 - \frac{1}{4}\left(\Sigma-\frac{2}{3}\Theta\right)^2,
\end{equation}
with the associated directional derivatives along $u^a$ and $e^a$ given by
\begin{eqnarray}
\dot{K} &=& \left(\Sigma - \frac{2}{3}\Theta\right)K, \\
\label{KDothat}
\quad\quad\quad \hat{K} &=& -\phi K.
\end{eqnarray}
The Misner-Sharp mass ($\mathcal{M}$) \cite{misnersharp} is a measure of the amount of gravitational mass within a given spherical shell at any instant of time. This is a purely geometrical measure, defined locally, in terms of the area radius of the spherical shell ($R$) in the following way \cite{exact}
\begin{equation}\label{M1}
\mathcal{M} =\frac{R}{2}\left(1 - \nabla_{a}R \nabla^{a}R\right)\,.
\end{equation}
The above can be written in terms of the Gaussian curvature of the shell $K \equiv R^{-2}$ as
\begin{equation} \label{M2}
\mathcal{M} =\frac{1}{2\sqrt{K}}\left(1 - \frac{1}{4K^3} \nabla_{a}K \nabla^{a}K\right)\,.
\end{equation}
We know that the symmetry of the spacetime dictates that the 4-gradient of any scalar $\Phi$ can be written as 
\begin{equation}
\nabla_a\Phi\equiv -\dot\Phi u^a +\hat\Phi e^a.
\end{equation} 
Substituting \eqref{K} and \eqref{KDothat} into \eqref{M2}, we finally get the expression for the Misner-Sharp mass in terms of the covariant scalars
\begin{equation}
\label{M3}
\mathcal{M} = \frac{1}{2K^{\frac{3}{2}}}\left(\frac{1}{3}\mu - \mathcal{E} - \frac{1}{2}\Pi\right).
\end{equation}
 We note that the above equation for the gravitational mass is an equivalent alternative representation without using the spatial integral of the component $T^0_{~~0}$ of the matter energy momentum tensor $T^a_{~~b}$. This representation is interesting as it transparently relates the total gravitational mass with matter inhomogeneity (generated by the tidal effects of the Weyl scalar) and matter anisotropy (generated by the anisotropic stress scalar $\Pi$ which is proportional to the difference between radial and tangential pressure).

 From the field equations \cite{clarkson2007} and equations \eqref{K}-\eqref{KDothat}, we can also write down the associated directional derivatives of this mass along $u^a$ and $e^a$ \cite{clarksonbarrett} as
\begin{eqnarray}
\label{Mdot}
\dot{\mathcal{M}} &=& \frac{1}{4K^{\frac{3}{2}}}\left[\left(\Sigma - \frac{2}{3}\Theta\right)\left(p + \Pi\right) -\phi Q\right], \\
\hat{\mathcal{M}} &=& \frac{1}{4K^{\frac{3}{2}}}\left[\phi\mu - \left(\Sigma-\frac{2}{3}\Theta\right)Q\right].
\end{eqnarray}

\section{Geometry of the Virial theorem} \label{sect4}

For a given spherical shell, let us now define
\begin{equation}
\alpha =  \frac{\mathcal{M}}{R} \equiv \mathcal{M}\sqrt{K}.
\end{equation}
If the shell is not trapped then the quantity $\alpha$ is strictly less than half. Now expanding \eqref{M2} and using the value of $K$ \eqref{K}, we obtain
\begin{equation}\label{vir1}
\phi^2 - \left(\Sigma - \frac{2}{3}\Theta\right)^2 = \frac{2\left(1-2\alpha\right)}{\alpha}\left[\frac{1}{3}\mu - \mathcal{E} - \frac{1}{2}\Pi\right].
\end{equation}
For the Buchdahl bound $\alpha=4/9$, we have 
\begin{equation}
 \frac{2\left(1-2\alpha\right)}{\alpha} =\frac{1}{2}. 
 \end{equation}
This is extremely interesting because in classical Newtonian mechanics, the Virial theorem states that for a system of particles interacting only through gravity in equilibrium, the total average kinetic energy of the particles of the system $\langle T \rangle$, and the total average potential energy $\langle V \rangle$, are related by
\begin{equation}\label{vir2}
\langle T \rangle=\frac{1}{2}\langle V \rangle \;.
\end{equation}
Hence we can write the geometric interpretation of the above expression in the form of a proposition. 
\begin{proposition}
We can identify the geometric notion of the `energy due to motion' with the volume expansion and shear of the matter flow lines as well as the spatial volume expansion of the preferred spatial congruence. On the other hand, the geometric counterpart of the `total potential energy' of the system is generated by the matter energy density and pressure anisotropy as well as the matter energy inhomogeneity by the Weyl scalar. Thus, we assert the following
\begin{eqnarray}\label{geom1}
\langle T \rangle_{geom} &=& \phi^2 - \left(\Sigma - \frac{2}{3}\Theta\right)^2, \\
\langle V \rangle_{geom} &=& \left[\frac{1}{3}\mu - \mathcal{E} - \frac{1}{2}\Pi\right].
\end{eqnarray}
\end{proposition}
We see that $\langle T \rangle_{geom} $ defined in \eqref{geom1} vanishes for $\alpha = \frac12$ according to \eqref{vir1}, and that gives the equation of the horizon in the $[u,e]$ plane. The fact that $\langle T \rangle_{geom} $ tends to zero at the black hole bound $\alpha=\tfrac12$ (by equation (\ref{geom1})), relates it directly to the norm of the 4-momenta of the infalling matter, which must vanish as the matter approaches the null horizon. Furthermore, since the Buchdahl bound is strictly the limiting case of a static, spherical and thermodynamically stable configuration in general relativity \cite{alho}, it can be stated that: 
\begin{proposition}
The geometric Virial stability is the limiting case of static, spherically symmetric and thermodynamically stable matter configurations.
\end{proposition}

Now let us ask an important question: ``{\em After attaining the state of Virial equilibrium, what are the conditions for a star to remain in that state despite continuous matter accretion?}" In other words, what happens when more matter accretes into the star? Is it possible that the star reaches the Virial equilibrium and then continues remaining in the Buchdahl star state instead of collapsing into a trapped region and forming a black hole? In the next section, we provide an explicit example of an accreting compact star in Virial equilibrium to answer this question. 

\section{An example: Accreting compact star in Virial equilibrium} \label{sect6}

Let us analyze the extra constraint 
 that the system must obey in order to continue accreting in the Virial equilibrium state, which is given by
\begin{eqnarray}\label{cons1}
C &:=& \langle T \rangle_{geom}-\frac{1}{2}\langle V \rangle_{geom} \nonumber\\
 &\equiv& \phi^2 - \left(\Sigma - \frac{2}{3}\Theta\right)^2+\frac{1}{2}\left[\mathcal{E} - \frac{1}{3}\mu + \frac{1}{2}\Pi\right]  = 0.
\end{eqnarray}
For this constraint to be satisfied for all time it should be consistently time propagated. In other words $C=0$ should imply $\dot{C}=0$. Therefore we take the directional derivative of the constraint along $u^a$ and use the following evolution equations derived from the Ricci and Bianchi identities \cite{clarkson2007} given by
\begin{eqnarray}
\label{phiDot}
\dot{\phi} &=& - \left(\Sigma-\frac{2}{3}\Theta\right)\left(\mathcal{A}-\frac{1}{2}\phi\right)+Q, \\
\label{SigmaDot}
\dot{\Sigma} - \frac{2}{3}\dot{\Theta} &=& -\mathcal{A}\phi + 2\left(\frac{1}{3}\Theta - \frac{1}{2}\Sigma\right)^2+ \frac{1}{3}\left(\mu + 3p\right) - \mathcal{E} + \frac{1}{2}\Pi, \\
\label{EDot}
\dot{\mathcal{E}} - \frac{1}{3}\dot{\mu} + \frac{1}{2}\dot{\Pi} &=& \left(\frac32\mathcal{E} + \frac{1}{4}\Pi\right)\left(\Sigma - \frac{2}{3}\Theta\right) + \frac{1}{2}\phi Q  - \frac{1}{2}\left(\mu + p\right)\left(\Sigma - \frac{2}{3}\Theta\right),
\end{eqnarray}
to calculate $\dot{C}$ and equate that to zero. 
We can then isolate an expression for the heat flux $Q$ (appearing in \eqref{phiDot}) and conclude that if the star continues to evolve as a Buchdahl star then it must expel  heat flux which is given by
\begin{equation}
\label{mainQ}
Q = \frac{\Sigma - \frac{2}{3}\Theta}{\phi}\left[\frac{1}{3}\left(\mu + 3p\right) + \frac{1}{2}\Pi - \mathcal{E}\right].
\end{equation}
There must be a continuous heat flux for \eqref{cons1} to remain conserved.

The necessary condition for a star that has attained the Virial equilibrium and for it to continue evolving in the same state of equilibrium is that the accreting matter has to expel the heat flux as Vaidya radiation in the exterior. 
Hence, it is natural to match the interior spacetime with an outgoing Vaidya exterior region (that describes a spherically symmetric spacetime with outgoing unpolarized radiation) \cite{deoliveira, fayos1, fayos2, fayos3, senovilla, fayos4, fayostorres} at the boundary of the accreting region with the area radius $R_{\mathcal{B}}$. Accordingly, the geometry of the exterior spacetime is described by the Vaidya metric given by
\begin{equation}
ds_2^2= -\left(1-\frac{2m(v)}{r_v}\right)dv^2-2dvdr_v+r_v^2(d\theta^2+\sin^2\theta d\phi^2).
\end{equation}
Here $v$ is the exploding null coordinate, $m(v)$ is the Vaidya mass function and $r_v$ is the Vaidya radius. The unit vectors along the timelike and preferred spacelike congruences for the Vaidya spacetime are given by
\begin{equation}
\tilde u^a=\left(1-\frac{2m(v)}{r_v}\right)^{-1/2}\left(\frac{\partial}{\partial v}\right)^a,
\end{equation}
and
\begin{eqnarray}
\tilde e^a&=&-\left(1-\frac{2m(v)}{r_v}\right)^{-1/2}\left(\frac{\partial}{\partial v}\right)^a+\left(1-\frac{2m(v)}{r_v}\right)^{1/2}\left(\frac{\partial}{\partial r_v}\right)^a,
\end{eqnarray}
where the tilde denotes the exterior spacetime. It is also possible to match the interior spacetime with a generalized Vaidya exterior at a comoving stellar boundary as seen in \cite{goswamijoshi}. 

The associated energy momentum tensor of a radiation fluid is
\begin{equation}
\tilde{T}_{ab} = \tilde{\mu} k_a k_b,
\end{equation}
where $k_a$ is a null vector $\left(k_a k^a = 0\right)$ and for spherical symmetry it is required to be $k_a = \tilde{u}_a + \tilde{e}_a$ representing outflowing radiation. The following constraints on the geometrical quantities, calculated according to the definitions appearing in section \ref{sect3}, arise
\begin{eqnarray}
&&\tilde\Sigma-\frac23\tilde\Theta = 0, \nonumber\\
&& \tilde\mu=3\tilde p=\tilde Q = \frac32\tilde\Pi,
\end{eqnarray}
given explicitly in \cite{betschart}.
Geometrically matching the first and the second fundamental forms across the boundary shell $\mathcal{B}$ of the collapsing star (as seen in \cite{pretty}), we get the following junction conditions \cite{santos, bonnor} given by
\begin{equation}\label{match1}
m(v)_{\mathcal{B}}=\mathcal{M}_{\mathcal{B}},
\end{equation}
\begin{equation}\label{match2}
[m(v)_{,v}]_{\mathcal{B}}= R^2_{\mathcal{B}}\; \left[\frac{\phi}{(\Sigma-\frac23\Theta)}\right]_{\mathcal{B}}\left[\left(\mu + 3p\right) + \frac{3}{2}\Pi - 3\mathcal{E}\right]_{\mathcal{B}},
\end{equation}
where the second condition is the pressure balance condition $\tilde{p}=Q$. In \eqref{match1}, $\mathcal{M}_{\mathcal{B}}$ is $\mathcal{M}$ in  \eqref{M3} evaluated at the boundary which is equal to the Vaidya mass $m(v)$ at the boundary. 

\section{Discussion} \label{sect7}
From the preceding calculations, we find that a star in the Virial equilibrium can remain in the same state under continuous accretion without collapsing into a black hole provided that it expels out the heat flux given by \eqref{mainQ} as Vaidya radiation in the exterior. That is, the heat generated by the tidal deformation of accreting matter has to be expelled out to retain the Virial equilibrium of the star. This, however, does not mean that no energy is accreted. What happens is that the mass to radius ratio $(\mathcal{M}+\delta \mathcal{M})/(R+\delta R)=4/9$ remains unchanged. However, this happens only in the special case when the constraint \eqref{cons1} is satisfied and the star escapes getting trapped to form a black hole. Hence, the question posed at the end of section \ref{sect4} is answered in the affirmative in this special case. On the other hand, in a general accretion process, this constraint will not be satisfied then the Virial equilibrium will be disturbed and the star can collapse into a black hole. 

It is however interesting to note that we have geometrically established the equivalence between the Buchdahl bound and the Virial equilibrium by identifying the geometric analogue of the kinetic energy and potential energy. The realization that gravitational energy, which is computed from the unique exterior metric, could play a role in governing how compact a star can be is indeed very interesting and revealing. A Buchdahl star is defined by the gravitational energy being half of the non-gravitational energy. In an effort to understand this characterization \cite{jcap2} one is led to the Virial theorem.

Stellar equilibrium is maintained by hydrostatic pressure counteracting gravity. What would be the limiting case of this balance of forces? Could it simply be a particle configuration without fluid structure where gravity is counterbalanced by the kinetic energy of particles in motion alone? This question is important as the Virial theorem is applicable for a system of particles in motion interacting only through gravity. Such a situation is offered by the Vlasov kinetic matter governed by the Einstein-Vlasov equations \cite{And-Rein06, And11}. It is conceivable that as a star reaches the limiting state of compactness, its interior may not be able to sustain fluid structure and it breaks into free elements interacting only through gravity. Then internal energy (gravitational energy) would be the measure of average kinetic energy and similarly, non-gravitational energy of average potential energy. This is how the Virial theorem governing the equilibrium of a Buchdahl star emerged.

What could be the limiting fluid configuration for the limiting compactness? Appealing to common sense, it should be the one which cannot be further compressed. That is, an incompressible fluid of constant density which is in fact described by the unique Schwarzschild interior solution of the Einstein equations, and it is matched to the Schwarzschild exterior vacuum at the boundary. Then it turns out that the requirement of pressure at the centre remaining bounded leads to the Buchdahl bound, $\mathcal{M}/R \leq 4/9$. So for the Buchdahl star where the bound is saturated, the central pressure becomes infinite. Moreover, the constant density fluid makes sound velocity infinite which cannot be physically accepted. Thus, the saturation of the Buchdahl bound defining the Buchdahl star cannot be reached by a fluid distribution and therefore, one has to look for alternative distributions \cite{jcap2}. An example of this could be free particles in motion interacting only through gravity, referring to the Vlasov kinematic matter. Then its equilibrium is governed by the Virial theorem. 

Conclusively, Vlasov kinetic matter seems to be the most promising matter source for the Buchdahl star interior. It would therefore be very pertinent to solve the Einstein-Vlasov equations \cite{And-Rein06, And11} for the Buchdahl star interior. It may also be noted that the velocity of particles at the Buchdahl surface is $v^2=8/9$ \cite{sum-dad22} while for the black hole, $v^2=1$. It could therefore be imagined that free particles in the Buchdahl interior are moving with $v^2=8/9$ and for the black hole interior with $v^2=1$! This is an interesting and physically insightful way of understanding the two most compact objects in the Universe, one with a horizon (black hole) and the other without (Buchdahl star).

It is very illuminating and aesthetically satisfying that the limiting compactness is attained when particles interact only through gravity and nothing else. This is the most general and elemental consideration and hence, it is pertinent to seek its purely geometric realization. That is precisely the aim and purpose of this paper. We provide the geometric version of the Virial theorem where kinetic and potential energy are defined in terms of geometric scalars. This new and novel perspective is indeed very remarkable as it makes no reference to the interior matter models at all.

\begin{acknowledgements}
ND wishes to thank the University of KwaZulu-Natal (UKZN) for a visit which has facilitated this collaboration. RG and CH are supported by the National Research Foundation (South Africa) and UKZN.
\end{acknowledgements}


\begin{thebibliography}{99}
	
\bibitem{chandra} S. Chandrasekhar, {\em Principles of Stellar Dynamics} (Chicago, IL: University of Chicago Press) (1942).
\bibitem{buch1} H. A. Buchdahl, \textit{Phys. Rev.} {\bf 116}, 1027 (1959).

\bibitem{buch2} H. A. Buchdahl, \textit{Astrophys. J.} {\bf 146}, 275 (1966).
\bibitem{bondi} H. Bondi, \textit{Mon. Not. R. Astron. Soc.} {\bf 282}, 303 (1964).
\bibitem{islam} J. N. Islam,  \textit{Mon. Not. R. Astron. Soc.} {\bf 145}, 21 (1969).
\bibitem{wald} R. M. Wald, {\em General Relativity} (Chicago, IL: University of Chicago Press) (1984).
\bibitem{stephani} H. Stephani, {\em Relativity} (Cambridge, UK: Cambridge University Press) (2004).
\bibitem{and1} H. Andreasson, \textit{J. Diff. Equations} {\bf 245}, 2243 (2008).
\bibitem{kar} P. Karageorgis and J. Stalker, \textit{Class. Quant. Grav.} {\bf 25}, 195021 (2008).

\bibitem{jcap1} N. Dadhich, \textit{JCAP} {\bf 04}, 035 (2020).
\bibitem{bro-yor} J. D. Brown and J. W. York, \textit{Phys. Rev. D} {\bf 47}, 1407 (1993).
\bibitem{jcap2} N. Dadhich, arxiv:2212.06745 (2022).

\bibitem{And11} H. Andreasson,  \textit{Living Rev. Relativ.} {\bf 14}, 4 (2011).
\bibitem{And-Rein06} H. Andreasson and G. Rein, \textit{Class. Quant. Grav.} {\bf 24}, 1809 (2007).

\bibitem{maxforce} N. Dadhich, \textit{Phys. Rev. D} {\bf 105}, 064044 (2022).


\bibitem{clarksonbarrett} C. A. Clarkson and R. K. Barrett, \textit{Class. Quantum Grav.} {\bf 20}, 3855 (2003).
\bibitem{clarkson2007} C. Clarkson, \textit{Phys. Rev. D} {\bf 76}, 104034 (2007).
\bibitem{goswamiellis} R. Goswami and G. F. R. Ellis, \textit{Class. Quantum Grav.} {\bf 38}, 085023 (2021).
\bibitem{misnersharp} C. W. Misner and D. H. Sharp, \textit{Phys. Rev. D} {\bf 136}, B571 (1964).
\bibitem{exact} H. Stephani, E. Herlt, M. MacCullum, C. Hoenselaers and D. Kramer, {\it Exact Solutions to Einstein’s Field Equations} (Cambridge, UK: Cambridge University Press) (2003).

\bibitem{alho} A. Alho, J. Nat\'ario, P. Pani and G. Raposo, \textit{Phys. Rev. D} \textbf{106}, L041502 (2022).

\bibitem{deoliveira} A. K. G. De Oliveira, N. O. Santos and C. A. Kolassis, \textit{Mon. Not. R. Astron. Soc.} \textbf{216}, 1001 (1985).
\bibitem{fayos1} F. Fayos, X. Ja\'en, E. Llanta and J. M. M. Senovilla, \textit{Class. Quantum Gravit.} \textbf{8}, 2057 (1991).
\bibitem{fayos2} F. Fayos, X. Ja\'en, E. Llanta and J. M. M. Senovilla, \textit{Phys. Rev. D} \textbf{45}, 2732 (1992).
\bibitem{fayos3} F. Fayos, J. M. M. Senovilla and R. Torres, \textit{Phys. Rev. D} \textbf{54}, 4862 (1996).
\bibitem{senovilla} J. M. M. Senovilla, \textit{Gen. Relativ. Gravit.} \textbf{30}, 701 (1998).
\bibitem{fayos4} F. Fayos, J. M. M. Senovilla and R. Torres, \textit{Class.  Quantum Grav.} \textbf{20}, 2579 (2003).
\bibitem{fayostorres} F. Fayos and R. Torres, \textit{Class. Quantum Grav.} \textbf{21}, 1351 (2004).
\bibitem{goswamijoshi} R. Goswami and P. S. Joshi, \textit{Phys. Rev. D} \textbf{76}, 084026 (2007).

\bibitem{betschart} G. Betschart and C. A. Clarkson, \textit{Class. Quantum Grav.} \textbf{21}, 5587 (2004).


\bibitem{pretty} P. N. Khambule, R. Goswami and S. D. Maharaj, \textit{Class. Quantum Grav.} {\bf 38}, 075006 (2021).

\bibitem{santos} N. O. Santos, \textit{Mon. Not. R. Astron. Soc.} \textbf{216}, 403 (1985).
\bibitem{bonnor} W. B. Bonnor, A. K. G. De Oliveira and N. O. Santos, \textit{Phys. Rep.} \textbf{181}, 269 (1989).


 \bibitem{sum-dad22} S. Chakraborty and N. Dadhich, \textit{Eur. Phys. J. C} \textbf{83}, 677 (2023).

	 
\end{thebibliography}
\end{document}